\documentstyle[twocolumn,aps,epsfig,floats]{revtex}

\newcommand{\be}{\begin{equation}}
\newcommand{\ee}{\end{equation}}
\newcommand{\bea}{\begin{eqnarray}}
\newcommand{\eea}{\end{eqnarray}}

\begin{document}

\twocolumn[\hsize\textwidth\columnwidth\hsize\csname @twocolumnfalse\endcsname

\title {Local Defect in Metallic Quantum  Critical Systems}
\author{A. J. Millis$^{a}$, D. K. Morr$^{b}$, and J. Schmalian$^{c}$}
\address{$^{a}$ Center for Materials Theory, Department of Physics and
Astronomy, Rutgers University, Piscataway, NJ 08854\\ $^b$
Theoretical Division, Los Alamos National Laboratory, Los Alamos,
NM 87545\\ $^c$ Department of Physics and Astronomy and Ames
Laboratory, Iowa  State University, Ames, IA 50011}
\date{\today}
\draft
\maketitle
\begin{abstract}
 We present a theory of a single point, line or plane defect coupling to the
 square of the order parameter in a metallic system near
 a quantum critical point  at or above its upper critical dimension.
  At criticality, a spin droplet is nucleated around the
 defect with droplet core size determined by the
strength of the defect potential. Outside the core a   universal slowly
 decaying tail of the droplet is found,  leading to many dissipative channels
coupling to the droplet and  to a complete suppression of quantum tunneling.
 We propose an NMR experiment to measure the impurity-induced
changes in the local spin susceptibility.
\end{abstract}
 ] \narrowtext The behavior of
'droplets' of local order in a non-ordered background is an issue of wide
relevance in condensed matter physics. One particularly interesting
sub-class of problems concerns 'droplets' induced by defects in nearly
critical systems. A long-standing problem in heavy fermion physics concerns
the very small magnetic moments which have been observed in several
materials \cite{Broholm88,Amato98} and may be related to grain boundaries
and other structural defects\cite{Aronson97,Halperin98,Luke98}. In a
colossal magnetoresistance material, magnetic order was observed to be
enhanced near grain boundaries.\cite{Aeppli99} A related issue is the
magnetism induced in high temperature superconductors by apparently
non-magnetic substituents such as ${\rm Zn}$\cite{Alloul92}, which have been
interpreted\cite{Slichter99,Morr00} as spin droplets induced in a nearly
critical system (although other interpretations exist also\cite{Lee94}).
Nucleation of regions of charge density wave order around defect sites on
the surface of a 'correlated' material was reported by\cite{Weitering99}.
'Quantum Griffiths' effects and 'Kondo disorder' are presently of intense
interest.\cite{Maclaughlin96,Neto98,Neto00,Narayanan00}. The problem bears
on the fundamental issue of the Kondo effect near a quantum critical point
\cite{Larkin72}. Finally, recording of information involves the polarization
of small domains, whose long time dynamics and stability are of great
importance.

This Letter presents the theory of the local polarization
('droplet') induced by a single defect in an otherwise
nondisordered system which is near a quantum critical point at or
above it upper critical dimension, $d_{{\rm u}}$. These
restrictions allow a controlled theoretical treatment and apply to
a wide range of systems including metallic magnets in dimensions
$d=2,3$ \cite{Hertz76,Millis94} and ``quantum paraelectric'' (i.e.
nearly ferroelectric) systems in $d=3$\cite{Mueller91}. We study
defects which
couple to the square of the order parameter, i.e. change the 'local $T_{{\rm %
c}}$', and thus create small regions where order is more favored
than in the pure system. We address three questions: under which
circumstances does the defect create a 'droplet', a region about
the defect in which the order parameter is non-vanishing (at least
on short time scales)? What is the size and general properties
of the droplet? What are the relevant fluctuations? Our work is
complementary to that of Vojta and Sachdev \cite {Sachdev99}, who
studied a linear coupling of the defect to the order parameter in
a quantum critical system below $d_{{\rm u}}$. It is also related
to the work of Castro-Neto and Jones \cite{Neto98,Neto00}, but
differs in several important aspects.

Our starting point is a quantum Ginzburg-Landau action for an order
parameter field $\phi $. After obtaining its mean field solution  we
consider fluctuation corrections, which are tractable because the dimension
of the system is above $d_{{\rm u}}$.
Halperin and Varma used a similar approach for a classical
system \cite{Halperin76}. Our action, in
conveniently scaled variables, is, in space and imaginary time:
\begin{eqnarray}
{\cal S} &=&{\cal S}_{{\rm dyn}}+\frac{1}{2}\int_{0}^{E_{0}/T}d\tau \int
d^{d}x\left\{ \left( V\left( {\bf x}\right) +\kappa ^{2}\right) \text{ }\phi
({\bf x,}\tau )^{2}\text{ }\right.   \nonumber \\
&&+\left. \left( \nabla \phi ({\bf x,}\tau )\right) ^{2}+\frac{1}{2}\phi (%
{\bf x,}\tau )^{4}\right\} .  \label{S}
\end{eqnarray}
Here, $\kappa $ determines the distance of the bulk system to the
critical point. We measure lengths in units of the bare
correlation length of the problem (typically of the order of a
lattice constant) and energies in terms of the condensation
energy\ $E_{0}$ obtained by evaluating the static, spatially
uniform free energy with $\kappa =1$. In the following, we
consider only symmetrical defects, which are characterized by a
dimensionality $d_{d}$ (i.e., the number of dimensions along which
the defect potential, $V$, remains constant), a length scale $a$
(expected to be of the order of a lattice constant) over which $V$
decays in the $D=d-d_{d}$ transverse directions, and a
dimensionless strength $v=-\int d^{D}rV({\bf r})$ ($v>0$
corresponds to a local tendency towards order). Here, ${\bf r}$
refers to the $D=d-d_{d}$ transverse components of the
$d$-dimensional vector ${\bf x}$, where $d_{d}=0,1,2$ corresponds
to a point, line and plane defect, respectively.

The dynamic term ${\cal S}_{{\rm dyn}}$ takes the general form
\begin{equation}
{\cal S}_{{\rm dyn}}=\frac{T}{2}\sum_{{\bf q},\omega _{n}}\left( \frac{1}{c_{%
{\bf q}}^{2}}+\frac{1}{\Gamma _{{\bf q}}|\omega _{n}|}\right) |\omega
_{n}\phi ({\bf q},\omega _{n})|^{2}.  \label{Sdyn}
\end{equation}
where the coefficients $c$ and $\Gamma $ depend on whether the
system is overdamped (metallic case) or not, on the symmetry of
the order parameter, and on whether it is conserved. Examples
include: (1) the undamped Ising antiferromagnet, with $\Gamma
^{-1}=0$ and $c=const$; (2) the undamped Ising ferromagnet with a
conserved order parameter, $\Gamma ^{-1}=0$ and $c \sim 1/q$; (3)
the metallic (overdamped) antiferromagnet \cite {Hertz76,Millis94}
$\Gamma =const$; (4) the metallic Ising ferromagnet
$\Gamma _{{\bf q}}\sim q$. The form of ${\cal S}_{{\rm dyn}}$
combined with the static part of the free energy defines a
(mean-field) dynamic exponent $z$ which is $z=1,2,2,3$
respectively for the  cases listed above. The effective
dimensionality of the quantum phase transition problem defined by
$S$ is $d_{{\rm eff}}=d+z$ and we restrict to
$d_{{\rm eff}}\geq 4$.

We now sketch the essential features of the mean field solution (details
will be given elsewhere \cite{Millis01a}). We focus only on the transverse
dimensions and assume $\ V(r>a)=0$ so that for $r>a$ the mean field equation
is
\begin{equation}
-\nabla ^{2}\phi _{0}+\kappa ^{2}\phi _{0}+\phi _{0}^{3}=0.
\label{meanfield}
\end{equation}
For $0\leq \kappa \ll 1$, the solution is of the form $\phi
_{0}(r)=r_{0}^{-1}f(r/r_{0},\kappa r_{0})$ where $f$ is dimensionless. The
length scale $r_{0}$ is determined by connecting the solution of Eq.(\ref
{meanfield}) (i.e., for $r>a$) to the one for $r<a$ and thus depends on the
defect strength, $v$. For $\kappa r_{0}>1$ the $\phi ^{3}$ term may be
neglected at all $r$ and the solution is the familiar exponentially decaying
one. In the more interesting case $\kappa r_{0}<1$ and $D\leq 3$, the
behavior at $r<\kappa ^{-1}$ is controlled by the nonlinearity and the scale
$r_{0}$ defines the size of the droplet. We find (the $g_{D}$ are constants):

\bigskip

\begin{tabular}{|c||c|c|c|}
\hline
$D$ & $r<r_{0}$ & $r_{0}<r<\kappa ^{-1}$ & $\kappa ^{-1}<r$ \\ \hline\hline
$1$ & $g_{1}r_{0}^{-1}$ & $\sqrt{2}/r$ & $\sqrt{2}\kappa e^{-\kappa r}$ \\
\hline
$2$ & $\frac{g_{2}\ln \left( \frac{r}{r_{0}}\right) +g_{2}^{\prime }}{r_{0}}$
& $1/r$ & $e^{-\kappa r}/r^{1/2}$ \\ \hline
$3$ & $g_{3}$ & $\ln ^{-1/2}\left( \frac{r_{0}}{r}\right) /r$ & $e^{-\kappa
r}\ln ^{-1/2}\left( \kappa r_{0}\right) /r$ \\ \hline
$>3$ & $g_{D}r_{0}$ & $r_{0}^{D-3}/r^{D-2}$ & $e^{-\kappa r}r^{(1-D)/2}$ \\
\hline
\end{tabular}

\bigskip

In $D=1,2$, and in $D=3$ up to logarithms, 
the $r_{0}<$ $r<\kappa ^{-1}$ behavior of $\phi
_{0}\left( r\right)$ is independent of the short length scale
physics encoded in $r_{0}$. In all dimensions $\int d^{D}r \, \phi
_{0}(r)$ diverges at criticality and in $D\geq 2$ $\int d^{D}r \,
\phi _{0}^{2}(r)$ diverges. Thus many physical properties are
dominated by the exponential tail of $\phi _{0}(r)$. Fig. ~\ref{Fig1}a
shows a schematic
picture of the droplet amplitude.

%
%

\begin{figure} [t]
\begin{center}
\leavevmode
\epsfxsize=7cm
\epsffile{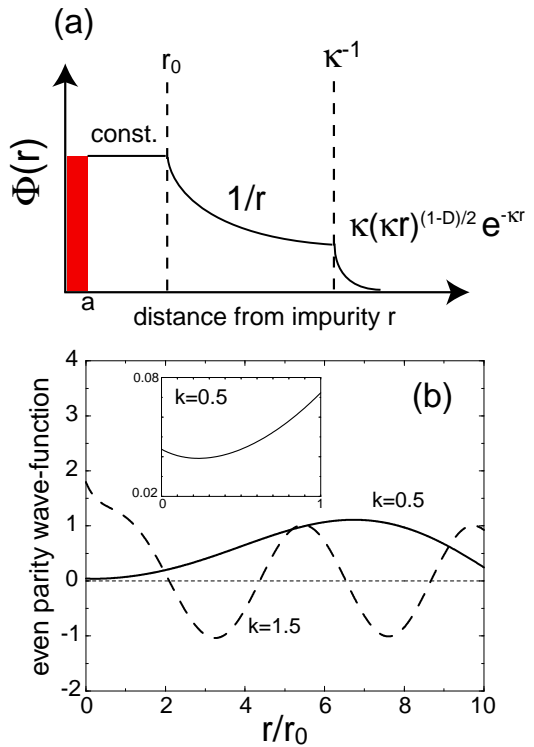}
\end{center}
\caption{ {\it (a)} Schematic $r$-dependence of the droplet
amplitude, $\phi_0(r)$. In $D=2$, $\phi_0(r)=const.$~for $r<r_0$
up to logarithmic corrections. {\it (b)} Dependence of even parity
wave-functions for  $k=0.5$ and $k=1.5$ on distance from the
defect. Inset: near-defect ($r \approx r_0$)
region for $k=0.5$. } \label{Fig1}
\end{figure}
The scale $r_{0}$ may be estimated by substituting our results for
$\phi _{0}(r)$ into Eq.(\ref{S}) and minimizing with respect to
$r_0$. This also determines the binding energy, $E_{{\rm bind}}$,
of the droplet which forms at temperatures $T<E_{{\rm bind}}$.
Alternative approaches including scaling, exact solutions ($D=1$),
and numerics give identical results \cite {Millis01a}.
In all cases the energetics are dominated by the droplet 'core'
region $r<r_{0}$. At criticality, an arbitrarily weak potential
induces a
droplet in $D=1,2$ but in $D=3$ a critical strength is required. In $D=1$ $%
r_{0}\sim v^{-1}$ and $E_{{\rm bind}}\sim -E_{0}v^{3}$ while in $D=2$ $%
r_{0}\sim e^{1/v}$ and $E_{{\rm bind}}\sim -E_{0}ve^{-2/v}$ In $D=3$ a
critical value $v^{\ast }\sim 1$ is required for droplet formation, and for $%
v>v^{\ast }$, $r_{0}$ is of order $a$ while $E_{{\rm bind}}\sim -E_{0}\frac{v%
}{v^{\ast }}(\frac{v}{v^{\ast }}-1)a^{-2}$. The droplet size cannot exceed $%
\kappa ^{-1}$, which yields an estimate for the critical potential $%
v^*(\kappa)$ $v>v^{\ast }=2\kappa $ ($D=1)$ and $v>v^{\ast }=-2\pi / \log
(\kappa a)$ ($D=2$). The droplet magnetization $M_d$ in the ferromagnetic
case is given by the integral of $\phi_{0}\left( {\bf r}\right) $ which
diverges as $\kappa ^{1-D}$ (logarithmically in $D=1$) as criticality is
approached. In an antiferromagnetic system with characteristic wave vector $%
{\bf Q}$, $M_{d} \sim \left| {\bf Q}\right| ^{1-D}$ which for $D\geq 1$ is a
number of order unity even at criticality. We show below that these mean
field results are not significantly affected by fluctuations, at least in
metallic systems.

Gaussian fluctuations may be treated by expanding to quadratic
order about the mean field solution, leading to the action:
\begin{equation}
{\cal S}_{{\rm GF}}={\cal S}_{{\rm dyn}}\ +\frac{T}{2}\int
d^{d}x\sum_{i\omega _{n}}\psi ({\bf x},i\omega _{n})\widehat{L}\ \psi \left(
{\bf x,}i\omega _{n}\right)   \label{effp}
\end{equation}
where $\widehat{L}=\kappa ^{2}+V\left( {\bf r}\right) +3\phi
_{0}^{2}\left( {\bf r}\right) -\nabla _{{\bf x}}^{2}$.
$\widehat{L}$ has only positive energy eigenfunctions. At
criticality all of these are extended but if $\kappa \not = 0$
then for $v$ in a small range above $v^{\ast}$, a bound state of
energy $0<E<\kappa^2$ may occur. The form of the potential (weak
slowly varying repulsion with attractive center) leads to
non-monotonic wave-functions with an upwards cusp at the defect
scale $a$, a decrease with distance in the range between $a$ and
$r_0$, and then (for extended states) an increase back to the unit
amplitude of a propagating plane wave. This $r-$dependence is
shown for two  even parity wave-functions of the Gaussian
fluctuations in Fig.~\ref{Fig1}b. The effect of Gaussian
fluctuations on the droplet size and shape may be computed in
terms of the difference between the eigenfunctions of
$\widehat{L}$ and of $\kappa ^{2}-\nabla ^{2}$ and are found
\cite{Millis01a} not to change the long distance behavior of the
droplet (as expected from conventional RG arguments
\cite{Hertz76,Millis94}).

A more important class of fluctuations changes the orientation
of the droplet.
In a compact ($d_{d}=0$) droplet these are rotations and 'instanton'
processes in which the droplet collapses and re-forms. In extended ($d_{d}>0$%
) droplets the important processes are motion of domain walls. Here, we 
sketch the results for a compact droplet; details and an analysis of the
moving domain wall case will be given elsewhere \cite{Millis01a}. The
dynamics of an isolated (not embedded in a critical system) droplet have
been previously studied \cite{Chudnovsky88}; the new feature 
here is the overdamped dynamics (in the metallic case).

We first consider a droplet with Ising symmetry, in which case the
instanton, i.e., the collapse of the droplet, is the important fluctuation
process. To estimate the action we substitute the ansatz $\phi ({\bf r},\tau
)=\phi _{0}({\bf r})\eta (\tau )$ into Eq.(\ref{S}) and retain leading time
derivatives. We find that the action corresponding to $2N$ instantons is:
\begin{eqnarray}
S_{2N} &=&\frac{\gamma }{2}\sum_{i\neq j=1...N}\log
(y_{i}-y_{j})(-1)^{i+j}+2N\left( \frac{y_{0}}{\zeta }+\frac{m}{y_{0}}\right)
\nonumber \\
&&+\frac{2N\gamma }{4}\int_{-1}^{1}du\int_{-1}^{1}dv\log
(1+y_{0}^{2}(u-v)^{2}),
\end{eqnarray}
with $\zeta =15E_{0}/(4E_{{\rm bind}})$, $m=E_{0}^{2}\sum_{{\bf q}}c_{{\bf %
q}}^{-2}\phi _{0}({\bf q})^{2}$ and $\gamma =E_{0}\sum_{{\bf q}}\Gamma _{%
{\bf q}}^{-1}\phi _{0}({\bf q})^{2}$. The terms proportional to $2N$ are the
action of a single instanton and $y_{0}$ (determined by minimizing the
second and third term of $S_{2N}$) is the duration of an instanton in units
of $E_{{\rm bind}}^{-1}$. Note that both $m$ and $\gamma $ diverge as $%
\kappa \rightarrow 0$ and that $\gamma =0$ in  undamped models.
In the weak dissipation ($\gamma \ll 1$) limit the standard macroscopic
quantum tunneling analysis \cite{LeggettRMP} leads to $S(y_{0})=8\sqrt{%
m/\zeta }+6m\gamma \zeta +{\cal O}\left( \gamma ^{2}\right)$ so the 'bare'
droplet tunneling rate is $\sim E_{0}e^{-S(y_{0})}$ and vanishes as
criticality is approached. Instanton-instanton interaction effects, which
arise from the first term in $S_{2N}$ are handled via a perturbative
renormalization group treatment and if $\gamma $ is less than a critical
value $\sim 1$ dissipative effects reduce the tunneling rate but not to zero.

In a metallic system near criticality, $\gamma \gg 1$, and the conventional
analysis does not apply. Our detailed results depend on the ratio $m/\gamma $%
. If $m/\gamma \gg 1$ (ferromagnetic case) then minimization leads to $%
S(y_{0})=2\gamma \left[ \log( m/2\gamma) +1\right]$ while for $m/\gamma \ll
1 $ we find $S(y_{0})=3m \left( \gamma/6m \right)^{1/3}$. In either case,
dissipation strongly suppresses the bare tunneling rate, and the large value
of $\gamma $ puts the action on the localized side of the Caldeira-Leggett
phase boundary, implying that tunneling processes are completely suppressed
on long time scales.

Our treatment closely parallels Hamann's formulation \cite{Hamann71} of the
Kondo dynamics of a single spin in a noncritical metal. Hamann found $%
S(y_{0})=\ln (1/JN_{0})$ ($J$ is the Kondo coupling and $N_{0}$ is
the Fermi surface density of states) and $\gamma =(1-JN_{0})^{2}$.
The crucial difference is that in our problem the large size of
the droplet allows many dissipative channels to couple to it,
leading to much stronger dissipative effects. Castro-Neto and
Jones \cite{Neto98} argued that the tunneling of a droplet could
be mapped onto a single-channel Kondo problem; in our model this
is not the case. A subsequent paper \cite{Neto00} considered a
droplet consisting of a large number of elementary $S=1/2$ spins
locked together by some magnetic interaction in a nearly critical
system neglecting the $1/r$ tail of the droplet and studying the effect
of dissipation on the bare tunneling rate. Their specific results
therefore differ from ours. Their crucial findings were that for
antiferromagnetic systems droplets as large as $10^{3}$ spins
could tunnel and that dissipative effects only become important below
an exponentially small scale, leaving a wide regime where 
quantum Griffiths behavior occurs, whereas
we find that near criticality dissipation always dominates and
quantum effects are  suppressed.

In droplets with ${\rm XY}$ or higher symmetry, rotational fluctuations must
be considered. We integrate out the conduction electrons of an action of the
type discussed in \cite{Schulz89} and assume that $\phi $ is characterized
by an amplitude $\phi _{0}({\bf r})$, obtained by solving Eq.(\ref{meanfield}%
), and a direction ${\bf n}(\tau )=\left( \cos (\theta (\tau )),\sin (\theta
(\tau )),0\right)$ specified by an angle $\theta$. Expanding in the angular
variables and retaining leading time derivatives gives
\begin{eqnarray}
S_{xy} &=&\frac{1}{2}\sum_{k,\omega }\chi _{zz}^{-1}(k,\omega
)\left| \phi _{z}(k,\omega )\right| ^{2} \nonumber \\
&&\hspace{-0.9cm} -\frac{\gamma_{xy}}{2}\int d\tau _{1}d\tau
_{2}\partial _{\tau } {\bf n(\tau } _{1})\cdot \partial _{\tau
}{\bf n(\tau }_{2})\ln \left( \frac{\tau _{0}^{2}+(\tau _{1}-\tau
_{2})^{2}}{\tau _{0}^{2}}\right) \nonumber \\ &&\hspace{-0.9cm}
-\frac{uM^{2}}{8}\int d\tau \left( i\partial _{\tau } {\bf n\times
n\cdot} \widehat{{\bf z}}+h_{z}(\tau )\right) ^{2} \label{Sxy}
\end{eqnarray}
with $\ M^{2}=\int d^{D}r\phi _{0}^{2}({\bf r})$ $h(\tau )=2\int
d^{d}r\phi _{0}(r)\delta \phi _{z}(r,\tau )/M^{2}$ and
$\gamma_{xy}=(2 \pi)^{-d} \int d^{d}q \left| \phi
_{0}(q)^{2}\right|/\Gamma _{q}.$ We thus obtain the action
expected for a rotor with a large moment of intertia ($M$) in a
dissipative environment, precessing in an effective magnetic field
caused by the background spin fluctuations. Ref.~\cite{Zwerger86}
indicates that in the large dissipation limit the subtleties
associated with spin quantization may be neglected. A
straightforward variational estimate then yields sub-diffusive
long-time behavior, corresponding to a divergent susceptibility.

The presence and fluctuations of the droplet
are in principle observable via NMR
measurements of the spin lattice relaxation rate, $T_{1}^{-1}(r)\sim T{\rm Im%
}\,\chi (r,\omega )/\omega $ and local Knight shift. There are two
different contributions: from changes, due to the droplet, in the
extended 'Gaussian' spin fluctuations, and from 
the presence and  tunneling of the droplet itself. The two
contributions have very different timescales and position dependences.
The Gaussian fluctuations have characteristic frequency 
$\omega \sim \kappa^z$ and give rise to the usual divergence
of the bulk relaxation rate as criticality is approached. This
contribution to $1/T_1T$ is suppressed near the defect (as shown for example
in Fig. 2) essentially because the droplet reduces the amplitude
of the low energy wave functions in the near-defect region (cf Fig. 1b).
The droplet tunnelling processes provide a contribution to 
$\chi^{''}(r,\omega)$ which is proportional to the square of the droplet
amplitude  but varies
on a much slower timescale, which moreover vanishes exponentially
as criticality is approached, so that 
these processes drop out of the NMR frequency window, appearing
instead as a broadening $\propto 1/r_0$ of the NMR spectrum.
Further details will be given elsewhere \cite{Millis01a}.

%
%

\begin{figure} [t]
\begin{center}
\leavevmode \epsfxsize=7cm \epsffile{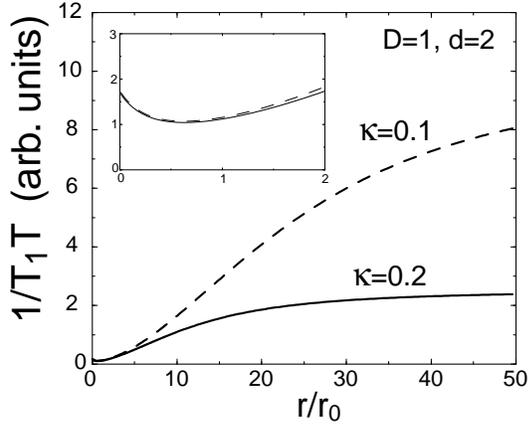}
\end{center}
\caption{Dependence of the Gaussian fluctuation contribution to
the  NMR relaxation rate on distance from the defect, calculated
for z=2, d=2 and D=1 and two distances from criticality. Inset:
expanded view of the near-defect ($r \approx r_0$) region.} \label{Fig2}
\end{figure}

In summary, we have presented a theory of a single defect 
in a quantum critical system at or above its upper critical dimension.
A crucial property is the $1/r$ 'tail' of the droplet extending 
into the surrounding medium.
The ease with which line and plane defects induce regions of local
order may be relevant to the small moments observed in heavy
fermion systems \cite{Broholm88,Amato98}. Our finding that near
criticality in a metallic system droplets behave in an essentially
classical manner leaves  no
significant parameter regime in which the quantum Griffiths
behavior discussed in \cite{Neto98,Neto00} exists. The strong
dimensionality dependence of our results has implications for the
general issue of Griffiths behavior near quantum criticality.
Finally, our results bear on the fundamental question of the Kondo
effect near a quantum critical point \cite{Larkin72}. A single
spin in a nearly critical system will similarly induce a large droplet,
which we believe will be prevented from
tunneling by dissipative effects .

DM acknowledges support by the U.S. DOE at the Los Alamos National
Laboratory, and AJM thanks D. Khmelnitskii and M.P.A. Fisher for helpful
conversations and acknowledges NSF DMR 00081075, and the British EPSRC,
Cambridge University, and the ITP at Santa Barbara
for support. JS acknowledges support by the Ames Laboratory operated for the
U.S. DOE by Iowa State University under contract No. W-7405-Eng-82.



\begin{references}
\bibitem{Broholm88}  G. Aeppli, E. Bucher, \ C. Broholm, J. K. Kjems, J.
Baumann, J. Hufnagl, Phys. Rev. Lett. {\bf 60}, 615 (1988).

\bibitem{Amato98}  A. De Visser et. al. J. Magn. Magn. Mater. {\bf 177-181},
435 (1998); A. Yaouanc et. al. preprint (cond-mat/0002011).

\bibitem{Aronson97}  M. C. Aronson t. al. PhysicaB {\bf 186-188}, 778
(1993); B. G. Demczyk, M. C. Aronson, B. R. Coles, and J. L. Smith, Philos.
Mag. Lett.{\bf 67}, 85 (1993).

\bibitem{Halperin98}  J. B. Kycia, J. I. Hong, M. J. Graf, J. A. Sauls, D.
N. Seidman, and W. P. Halperin, Phys. Rev. B {\bf 58}, R603 (1998).

\bibitem{Luke98}  G. M. Luke et. al. J. M. M. M. {\bf 177-181},
754 (1998).

\bibitem{Aeppli99}  Y.A. Soh, G. Aeppli, N.D. Mathur, and M.G. Blamire,
Phys. Rev. B {\bf 63}, 020402, (2001).

\bibitem{Alloul92}  H. Alloul, P. Mendels, H. Casalta, J. F. Marucco, J.
Arabski, Phys. Rev. Lett. {\bf 67}, 3140 (1991).

\bibitem{Slichter99}  J. Bobroff, et. al. Phys. Rev. Lett. {\bf 79}, 2117
(1997); {\em ibid }Phys. Rev. Lett. {\bf 80}, 3663 (1998).

\bibitem{Morr00}  D. K. Morr, J. Schmalian, R. Stern, C. P. Slichter, Phys.
Rev. Lett. {\bf 80}, 3662 (1998).

\bibitem{Lee94}  N. Nagaosa and P. A. Lee, Phys. Rev. Lett. {\bf 79}, 3755
(1997) and N. Nagaosa and T.-K. Ng, Phys. Rev. B {\bf 51}, 15588 (1997).

\bibitem{Weitering99}  H. H. Weitering et. al. Science {\bf 285}, 2107 (1999).

\bibitem{Maclaughlin96}  O. O. Bernal, D. E. MacLaughlin, H. G. Lukefahr, B.
D. Andraka, Phys. Rev. Lett. {\bf 75}, 2023 (1995); E. Miranda, V.
Dobrosavljevic, and G. Kotliar, Phys. Rev. Lett. {\bf 78}, 290 (1997), O.
Motrunich, S.-C. Mau, D. A. Huse, D. S. Fisher, Phys. Rev. B {\bf 61}, 1160
(2000).

\bibitem{Neto98}  A. H. Castro Neto, G. Castilla, B. A. Jones, Phys. Rev.
Lett. {\bf 81}, 3531 (1998).

\bibitem{Neto00}  A. H. Castro-Neto and B. A. Jones, Phys. Rev. {\bf B62}
14975 (2000).

\bibitem{Narayanan00}  R. Narayanan, T. Vojta, D. Belitz, and T. R.
Kirkpatrick, Phys. Rev. Lett. {\bf 82}, 5132 (1999).

\bibitem{Larkin72}  A. I. Larkin and V. I. Melnikov, Sov. Phys. JETP {\bf 61}
1231-42 (1972).


\bibitem{Hertz76}  J. A. Hertz, Phys. Rev. {\bf 14}, 1165 (1976).

\bibitem{Millis94}  A. J. Millis, Phys. Rev. B {\bf 48}, 7183 (1993).

\bibitem{Mueller91}  K. A. M\"{u}ller and \ H. Burkhard, Phys. Rev. B {\bf 19%
}, 3593 (1979).

\bibitem{Sachdev99}  S. Sachdev, C. Buragohain, and M. Vojta, Science {\bf %
286}, 2479 (1999); M. Vojta, C. Buragohain, and S. Sachdev, Phys. Rev. B
{\bf 61}, 15152 (2000).

\bibitem{Halperin76}  B. I. Halperin and C.M. Varma, Phys. Rev. B {\bf 14},
4030-44 (1976).

\bibitem{Millis01a}  A. J. Millis, D. K. Morr, and J. Schmalian
(unpublished).

\bibitem{Chudnovsky88}  E. M Chudnovsky and L Gunther, Phys. Rev.Lett {\bf 60%
} 661-4 (1988) and D. Loss, d. DiVincenzo and G.. Grinstein, Phys. Rev. Lett.%
{\bf 69 }3232-5 (1992).

\bibitem{LeggettRMP}  A. J. Leggett et. al. Rev. Mod. Phys. {\bf 59}, 1 (1987).

\bibitem{Hamann71}  D. R. Hamann Phys. Rev. {\bf B2}1373-92 (1970).

\bibitem{Schulz89}  H. J. Schulz, Phys. Rev. Lett. {\bf 65}, 2462 (1990).

\bibitem{Zwerger86}  W. Zwerger, A. T. Dorsey and M. P. A. Fisher, Phys.
Rev. {\bf B34} 6518-21 (1986).
\end{references}
\end{document}